# A Three State NDI Switch: Integration of Pendant Redox Unit for Conductance Tuning**

Dr. Yonghai Li,[#] Masoud Baghernejad,[#] Al-Galiby Qusiy,[#] Dr. David Zsolt Manrique,[#] Dr. Guanxin Zhang, Joseph Hamill, Dr. Yongchun Fu, PD Dr. Peter Broekmann, Prof. Wenjing Hong,* Prof. Thomas Wandlowski, Prof. Deqing Zhang,* Prof. Colin Lambert*

---

[*] Dr. Yonghai Li, Dr. Guanxin Zhang, Prof. Deqing Zhang
Organic Solids Laboratory, Institute of Chemistry, Chinese Academy of Sciences, Beijing 100190, China
dqzhang@iccas.ac.cn
Masoud Baghernejad, Joseph Hamill, Dr. Yongchun Fu, PD Dr. Peter Broekmann, Prof. Wenjing Hong, Prof. Thomas Wandlowski
Department of Chemistry and Biochemistry, University of Bern, Freiestrasse 3, CH-3012, Bern, Switzerland
hong@dcb.unibe.ch
Prof. Wenjing Hong
Department of Chemical and Biochemical Engineering, College of Chemistry and Chemical Engineering, Xiamen University, Xiamen 361005, P.R. China
Al-Galiby Qusiy, Dr. David Zsolt Manrique, Prof. Colin Lambert
Department of Physics, Lancaster University, Lancaster LA1 4YB, UK
c.lambert@lancaster.ac.uk
[#] These authors contributed equally

[**] This work was generously supported by National Natural Science Foundation of China and Ministry of Science and Technology (via 973 project), the Swiss National Science Foundation (200020-144471), the EC FP7 ITN "MOLESCO" project number 606728, UK EPSRC grants EP/K001507/1, EP/J014753/1, EP/H035818/1, and the University of Bern. A. Q. also acknowledged the financial support from Ministry of Higher Education and Scientific Research, Al Qadisiyah University, IRAQ.
Supporting information for this article is given via a link at the end of the document.

**Abstract:** We studied charge transport phenomena through a core substituted naphthalenediimide (NDI) single-molecule junctions using the electrochemical STM-based break junction technique in combination with DFT calculations. The conductance switch among three well-defined states is acquired by electrochemically controlling the redox state of the pendent diimide unit of the molecule in ionic liquid, and the conductance difference is more than one order of magnitude between di-anion states and neutral state. The potential dependent charge transport characteristics of the NDI molecules are confirmed by DFT calculations accounting for electrochemical double-layer effects on the conductance of the NDI junctions. This work suggests that the integration of redox unit in the pendent position with strong coupling to molecular backbone can significantly tune the charge transport of the single-molecule device by controlling different redox states.

Functional molecules with bi-/multi-stable states have intensively been studied[1] because they are potentially interesting building blocks for molecular-level electronics[2] and ultra-high density storage devices.[3] Various switchable molecules were fabricated by incorporating photochromic or redox-active moieties within the molecular charge transport pathway, e. g., tetrathiafulvalene (TTF),[4] benzodifuran (BDF),[5] anthraquinone (AQ),[6] and ferroncе (Fc).[7] In these studies, functional unit was involved in the charge transport pathway, which is expected to provide more significant tuning of single-molecule conductance. On the other hand, single-molecule devices with a pendant functional unit (Figure 1a), which is not directly involved in the charge transport pathway, are also of great interest because it offers much more flexibility for molecular design and synthesis, and therefore, finer tuning of the charge transport through the single molecule device.

Naphthalene diimide (NDI) have attracted much attention in organic electronics[8] and supramolecular chemistry[9] community acting as an electron acceptor with n-type semiconductor characteristics. More interestingly, substitution on the naphthalene core provides the opportunity to study charge transport through the naphthalene unit (marked as gray in Figure 1b) while diimide unit (marked as green in Figure 1b) served as pendant functional unit with strong coupling to the naphthalene backbone and response to external stimulus for instance, applied electrochemical gating. Varying applied potential to the NDI molecules, a sequence of two sequential electron transfer reactions transforms the neutral species (NDI-N) into the corresponding radical-anion (NDI-R), and finally into the di-anion species (NDI-D). Therefore this core substituted NDI molecule can be considered as a prototypical molecular junction to evaluate the effect of pendent redox groups on the single-molecule conductance.

In this communication, we report an electrochemically controlled STM-based break junction study[6] of a NDI-BT molecule with pendent diimide redox unit strongly coupled to the molecular backbone. We demonstrate that these reversible redox-transitions in the pendent diimide unit indeed cause pronounced changes in the charge transport through the single molecular NDI-BT junction. To get better understanding of the microscopic mechanism of the electrochemical gating, we employed density functional theory (DFT) to model the charge double layer in the molecular junction comprised of the ions in the supporting electrolyte and computed electrical conductance with non-equilibrium Green's function (NEGF) method as a function of the NDI-BT redox-states.

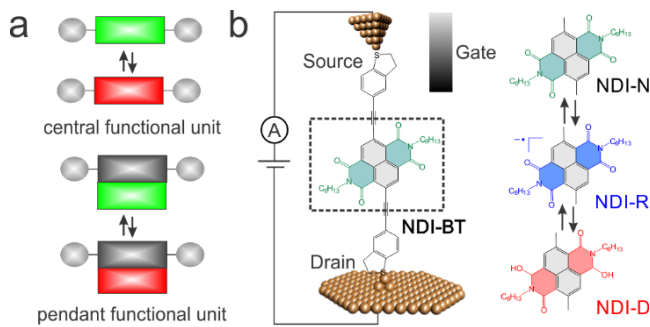

**Figure 1.** (a) Schematic illustration of single-molecule device with central redox unit in the charge transport pathway and pendant redox unit (b) Schematic illustration of electrochemically gated break junction experiment and molecular structure of NDI-BT in neutral state (NDI-N), radical anion state (NDI-R) and di-anion state (NDI-D).

The design of NDI-BT is based on the following considerations: i) NDI can be reversibly transformed into the respective radical anion and di-anion. Moreover, the NDI reduction potentials can be fine-tuned by chemical modification at its bay-positions; ii) the presence of adihydrobenzo[b]thiophene anchor group enables NDI-BT to be contacted to source-drain electrodes (Figure 1b);[10] iii) the presence of triple carbon-carbon bonds introduces a certain rigidity into the NDI-BT (detailed information on the synthesis and chemical characterization of the NDI-BT is provided in SI). In order to demonstrate the three-accessible redox states of the molecule, UV-Vis absorption spectra of NDI-BT were measured before and after addition of various amounts of hydrazine (reducing reagent). These three significantly different UV-Vis absorption spectra (See Figure S2) indeed prove the existence of three stable NDI-BT redox-states with distinguishable energy gaps.

The NDI-BT assembly was prepared on a Au(111) substrate by drop-casting of 30 µL of 0.5 mM NDI-BT in THF. As the complete reduction of the NDI-BT to the respective di-anion requires rather negative potentials, all cyclic voltammetry (CV) and STMBJ measurements were carried out in $HMImPF_6$ as supporting electrolyte in an oxygen-free environment. Compared to the electrochemical response of the bare Au(111)/$HMImPF_6$ interface (grey curve in Figure 2a), two pairs of reversible redox-peaks appear in the voltammogram when the NDI-BT layer is present (black curve in Figure 2a). The first redox peak is located around -0.85V vs Fc/Fc$^+$ and is attributed to the NDI-N/NDI-R redox process. The second redox peak is observed around -1.15V vs Fc/Fc$^+$ corresponds to the redox process of NDI-R to the NDI-D states. The voltammogram peaks are in good agreement with the measured CVs in $CH_2Cl_2$ (see Figure S3).

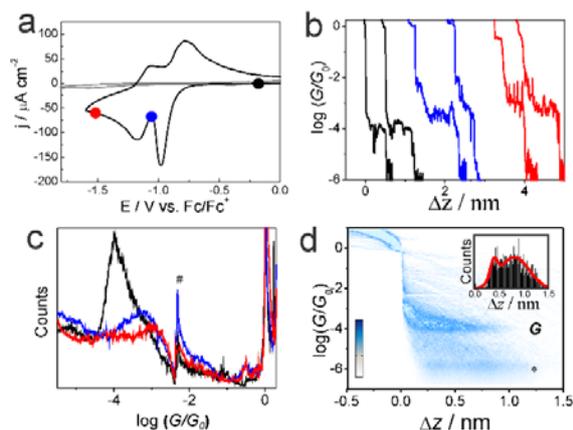

**Figure 2.** Figure Caption. Figure 2 (a) CVs of NDI-BT immobilized on the Au(111) electrode (black curve) and bare Au(111) (gray curve) in HMImPF6. The scan rate was 50 mV s$^{-1}$. (b) Typical conductance-distance traces at different potential black: NDI-N at -0.2 V, blue: NDI-R at -1.05 V, red: NDI-D -1.5 V versus Fc/Fc+ (c) conductance histogram constructed from 1000 individual traces sampled at different potential, # presents an artifact caused by the switching between different current measured range of the linear amplifier (d) 2D conductance-distance histogram of NDI-N, *presents the current measurement noise level during the electrochemical-STM BJ measurement.

The charge transport properties of the single NDI-BT molecule was studied by STM-BJ measurements at room temperature. For further technical details we refer to our previous work.[11] Figure 2b displays representative conductance (G) versus distance (Δz) traces measured at three different electrode potentials corresponding to NDI-N (black), NDI-R(blue), and NDI-D(red). After the contact formation between the Au tip and the Au substrate, the tip was withdrawn at a rate of 87 nm s$^{-1}$. All curves show an initial conductance feature at $G_0$ with $G_0 = 2e^2/h$ = 77.5 µS corresponding to the single gold-gold conductance[12]. Subsequently, the conductance abruptly decreases ("jump out of contact"[13]) by several orders of magnitude. The plateaus observed in the range form ~ $10^{-3}$ $G_0$ to $10^{-4}$ $G_0$ are assigned to the conductance features of the single molecular NDI-junctions. These strongly depend on the applied electrode (gate) potential.

To provide a more quantitative comparison among the conductance values of all three redox/conductance states, 1000 individual conductance traces were sampled for each potential. In order to extract statistically significant results these data were plotted in point histograms without any data selection. It is found that the conductance increased from $10^{-4}$ $G_0$ to $10^{-3.3}$ $G_0$ when the potential was swept from -0.2 V (NDI-N state) to -1.05 V vs (Fc/Fc$^+$ NDI-R state). Moreover, NDI-D sampled at -1.5 V vs. Fc/Fc$^+$ provided an even higher conductance of ~$10^{-3.0}$ $G_0$. The ON/OFF ratio from the NDI-N state to NDI-D state is of the conductance difference of around an order of magnitude, which is slightly larger than the previous studies on benzodifuran (BDF) and multiple states TTF switches,[4] and comparable with our recent single-molecule switch using an anthraquinone redox unit.[6]

Figure 2d shows the two-dimensional (2D) conductance vs. displacement histogram constructed from all data points.[14] The 2D conductance histogram provides direct access to the evolution of molecular junctions during the formation, stretching and break-down steps. In this histogram, the corresponding gold-gold breaking junction and additional high-density data cloud around $10^{-4.5}$ $G_0$ ≤ G ≤ $10^{-3.5}$ $G_0$, cantered on $10^{-4}$ $G_0$ are observed. This region represents the conductance range of the single NDI-N molecule bridging the gold electrodes. The relative displacement ($\Delta z$) histograms of the molecular junction evolution upon stretching was constructed by calculating the displacement from the relative zero position at 0.7 $G_0$ up to the end of the molecular conductance region of every individual trace.[15] Figure 2d inset displays the characteristic displacement histograms of NDI-N with a well-defined distribution at $\Delta z^*$ = (0.90 nm) as a measure of the most-probable plateau lengths of NDI-N molecular junctions. As observed, two peaks corresponding to pure tunnelling and molecular junction in figure 2D led to a junction formation probability of ~80 %.

The reversibility of the switching process was also evaluated by continuous conductance switching between different charge states. As shown in Figure 3, each conductance histogram are constructed from 1000 individual traces, and the applied potential changed per 1000 traces between -0.2 V (NDI-N), -1.05 V (NDI-R), and -1.5 V (NDI-D) versus Fc/Fc$^+$, respectively. It is found that the molecule can be switched forward and backward from NDI-N via NDI-R to NDI-D state (1st-5000th traces), or switching between NDI-N and NDI-R state (5000th to 7000th traces). These switching cycles suggest that the three charge states of the NDI molecule can be tuned reversibly by changing the applied potential. The break junction experiment returned to the initial NDI-N state with still quite pronounced conductance feature (top histogram in Figure 3a) even after more than 7000 stretching cycles, and the conductance remain constant for each specific redox states among different switching cycles(figure 3b), which suggests a high reversibility and stability of the NDI-BT molecular switch. It should be noticed that the conductance peak became slight broader during the measurement, which may be attributed to desorption of NDI-BT molecules from the electrodes, especially at NDI-R and NDI-D states.

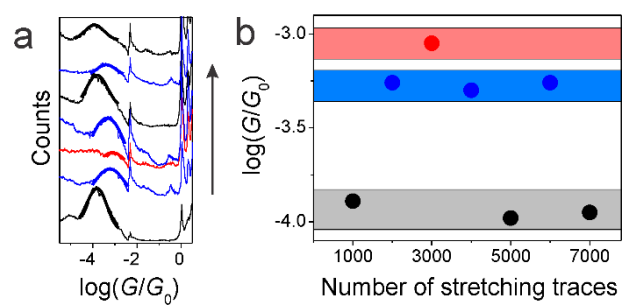

**Figure 3.** (a) Conductance histogram constructed from 1000 individual traces in each applied potential (b) Most-probable conductance value determined from conductance histograms shown in (a).

To understand the observed switching mechanism between the three states of electrochemical gating we have theoretically investigated the effect of the charge double layer on the conductance. The primary role of the charge double layer in our calculations is to control the number of electrons on the NDI-molecule. To perform conductance calculations we used the Gollum[16] non-equilibrium Green's function based quantum transport code and the optimal gold-molecule-gold junction geometries and the Hamiltonian matrix elements were obtained with SIESTA.[17] Further details of the calculations are in the SI. To accurately describe single occupied levels of the NDI all calculations were spin-polarized and the electron transmission coefficient function $T(E)$ is calculated as the average of the spin up and spin down transmission coefficients. The three stages of gating are modelled with the three types of charge double layers (CDLs) shown in Figure 4 and the detailed junction and double layer geometries are shown in Figure S10-S12 in the SI. To model the effect of gating, the distance (*Y*) between the double layers and the plane of the molecule is adjusted such that the number of extra electrons on the molecule is 0, 1 and 2 for the neutral, radical anion and di-anion stages, respectively. Figure 4a shows a negative charge double layer (the negative ions of the double layer lie closer to the plane of the NDI molecule) for which the NDI is neutral. Figures 4b and 4c show positive charge double layers that attract electrons from the gold leads to the molecule, creating the radical anion and di-anion states.

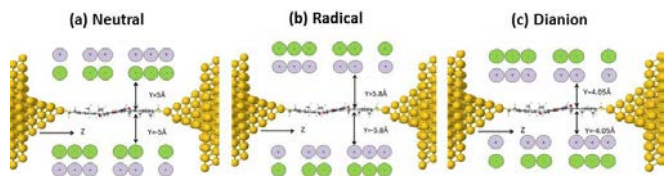

**Figure 4.** Junction geometries with charge double layers for the three states of electrochemical gating. (a) The neutral state with negative-positive CDL that adjusts the molecular charge to zero. (b) and (c) The radical anion and di-anion states with positive-negative CDLs.

To account for the fluctuating charge double layer structures in the experiments, the transmission coefficient and conductance were computed for different charge double layer arrangements, in which the anions and cations were randomly arranged at a fixed distance *Y*. These were then averaged to yield the theoretical conductance *G* and transmission coefficient *T(E)* shown in figure 5 (further details are in the SI). Figure 5a,b and c show the average transmission coefficients for the three states. The conductances, which are computed from the transmission coefficients at the Fermi energy using the room temperature Landauer formula, show an increasing trend moving from neutral state to the di-anion state. In the case of radical anion state the Fermi energy is trapped between two single electron resonances. This is apparent only in the spin-polarized calculation. The non-spin-polarized calculations give qualitatively different theoretical trends because single occupied orbitals are by definition located at the Fermi energy resulting in an unphysical high conductance for the radical state (see figure S7 in the SI). Figure 5d illustrates that the electrochemical gating experiment consistently increases the conductance as the molecular charge goes from the neutral state to the di-anion state. This trend agrees well with the theoretical charge double layer model as Figure 5d illustrates. The theoretical conductance values are higher than the measured ones for all three states, which can be attributed to the neglect of environmental and thermal effects.[18] DFT is also known to underestimate the HOMO-LUMO gap, resulting in an overestimated conductance.[13, 15, 19,20]

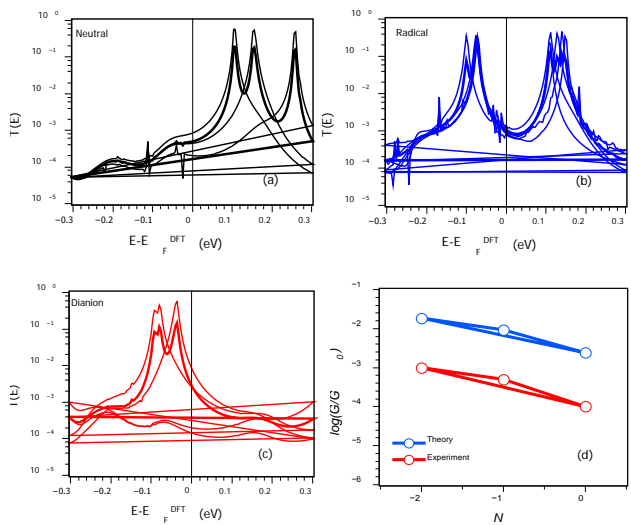

**Figure 5.** (a), (b) and (c) show the transmission curves for junction geometries with double layers located at distances *y=5Å, y=5.8Å* and *y=4.05Å*, respectively. Example junction geometries are shown in Figure 4 (a), (b) and (c). The continuous curves are the averaged transmission coefficients and the dotted curves show the transmission coefficients for different charge double layer arrangements. The colour code refers to the three different states, NDI-N (black), NDI-R (blue) and NDI-D (red). (d) shows the measured and computed averaged conductance values for the three states of the electrochemical gating experiment. The computed conductance values are calculated from the averaged transmission coefficient using the finite temperature Landauer formula (eq. 1 in the SI) with the temperature 300K.

To conclude, we have studied charge transport through a core substituted NDI single-molecule junctions using an STM-BJ technique under electrochemical gating and through ab initio simulations based on density functional theory. Benefiting from the wide potential window of the ionic liquid as a supporting electrolyte, we are able to access the three conductance states of the NDI molecule. These three conductance states showed well-separated conductance features with an ON/OFF ratio around one order of magnitude between the most conductive NDI-R and the least conductive NDI-N state. The switching process can be manipulated reversibly via the applied electrochemical gate potential. Agreement between theory and experiment has been demonstrated by introducing a newly developed charge double layer model. The tri-stable charge states of the NDI molecule provides a unique opportunity beyond bi-stable ON-OFF molecular switches, and is expected to lead to interesting logic gate and memory devices. More importantly, the comparable conductance changes between the NDI molecule with pendent diimide units and the molecules with redox units integrated in molecular backbones suggest that the pendent redox unit can also provide significant fine tuning of single-molecule conductance as internal gate unit of single-molecule device triggered by external stimuli, in this case, electrochemical gate. This offers a great flexibility for the molecular design and synthesis for future molecular electronics applications.

**Keywords:** single molecule • molecular electronics• naphthalene diimide • break junction• electrochemical gating

[1] a) A. C. Fahrenbach, C. J. Bruns, H. Li, A. Trabolsi, A. Coskun, J. F. Stoddart, *Accounts. Chem. Res.* **2014**, *47*, 482-493; b) M. Irie, T. Fulcaminato, K. Matsuda, S. Kobatake, *Chem. Rev.* **2014**, *114*, 12174-12277; c) B. L. Feringa, R. A. van Delden, N. Koumura, E. M. Geertsema, *Chem. Rev.* **2000**, *100*, 1789-1816; d) P. Ceroni, A. Credi, M. Venturi, *Electrochemistry of Functional Supramolecular Systems*, Wiley-VCH: Weinheim, Germany, **2010**.
[2] a) C. R. Arroyo, S. Tarkuc, R. Frisenda, J. S. Seldenthuis, C. H. M. Woerde, R. Eelkema, F. C. Grozema, H. S. J. van der Zant, *Angew. Chem. Int. Edit.* **2013**, *52*, 3152-3155; b) Z. Li, M. Smeu, S. Afsari, Y. Xing, M. A. Ratner, E. Borguet, *Angew. Chem. Int. Edit.* **2014**, *53*, 1098-1102; c) C. Huang, A. V. Rudnev, W. Hong, T. Wandlowski, *Chem. Soc. Rev.* **2015**, *44*, 889-901.
[3] a) H. Tian, *Angew. Chem. Int. Edit.* **2010**, *49*, 4710-4712; b) J. E. Green, J. W. Choi, A. Boukai, Y. Bunimovich, E. Johnston-Halperin, E. Delonno, Y. Luo, B. A. Sheriff, K. Xu, Y. S. Shin, H. R. Tseng, J. F. Stoddart, J. R. Heath, *Nature* **2007**, *445*, 414-417.
[4] a) N. J. Kay, S. J. Higgins, J. O. Jeppesen, E. Leary, J. Lycoops, J. Ulstrup, R. J. Nichols, *J. Am. Chem. Soc.* **2012**, *134*, 16817-16826; b) E. Leary, S. J. Higgins, H. van Zalinge, W. Haiss, R. J. Nichols, S. Nygaard, J. O. Jeppesen, J. Ulstrup, *J. Am. Chem. Soc.* **2008**, *130*, 12204-+.
[5] Z. Li, H. Li, S. Chen, T. Froehlich, C. Yi, C. Schönenberger, M. Calame, S. Decurtins, S.-X. Liu, E. Borguet, *J. Am. Chem. Soc.* **2014**.
[6] AK Ismael, I Grace, CJ Lambert, *Nanoscale* **2015** 7 (41), 17338-17342
[7] X.-S. Zhou, L. Liu, P. Fortgang, A.-S. Lefevre, A. Serra-Muns, N. Raouafi, C. Amatore, B.-W. Mao, E. Maisonhaute, B. Schoellhorn, *J. Am. Chem. Soc.* **2011**, *133*, 7509-7516.
[8] a) F. Wuerthner, M. Stolte, *Chem. Commun.* **2011**, *47*, 5109-5115; b) X. Zhan, A. Facchetti, S. Barlow, T. J. Marks, M. A. Ratner, M. R. Wasielewski, S. R. Marder, *Adv. Mater.* **2011**, *23*, 268-284; c) Z. Liu, G. Zhang, Z. Cai, X. Chen, H. Luo, Y. Li, J. Wang, D. Zhang, *Adv. Mater.* **2014**, *26*, 6965-6977.
[9] A. J. Avestro, D. M. Gardner, N. A. Vermeulen, E. A. Wilson, S. T. Schneebeli, A. C. Whalley, M. E. Belowich, R. Carmieli, M. R. Wasielewski, J. F. Stoddart, *Angew. Chem. Int. Edit.* **2014**, *53*, 4442-4449.
[10] P. Moreno-Garcia, M. Gulcur, D. Z. Manrique, T. Pope, W. Hong, V. Kaliginedi, C. Huang, A. S. Batsanov, M. R. Bryce, C. Lambert, T. Wandlowski, *J. Am. Chem. Soc.* **2013**, *135*, 12228-12240.
[11] C. Li, I. Pobelov, T. Wandlowski, A. Bagrets, A. Arnold, F. Evers, *J. Am. Chem. Soc.* **2007**, *130*, 318-326.
[12] B. Q. Xu, N. J. J. Tao, *Science* **2003**, *301*, 1221-1223.
[13] S. Y. Quek, M. Kamenetska, M. L. Steigerwald, H. J. Choi, S. G. Louie, M. S. Hybertsen, J. B. Neaton, L. Venkataraman, *Nat. Nanotech.* **2009**, *4*, 230-234.
[14] a) C. A. Martin, D. Ding, J. K. Sorensen, T. Bjornholm, J. M. van Ruitenbeek, H. S. J. van der Zant, *J. Am. Chem. Soc.* **2008**, *130*, 13198-13199; b) A. Mishchenko, L. A. Zotti, D. Vonlanthen, M. Buerkle, F. Pauly, J. Carlos Cuevas, M. Mayor, T. Wandlowski, *J. Am. Chem. Soc.* **2011**, *133*, 184-187.
[15] W. Hong, D. Z. Manrique, P. Moreno-Garcia, M. Gulcur, A. Mishchenko, C. J. Lambert, M. R. Bryce, T. Wandlowski, *J. Am. Chem. Soc.* **2012**, *134*, 2292-2304.
[16] J. Ferrer, C. J. Lambert, V. M. Garcia-Suarez, D. Z. Manrique, D. Visontai, L. Oroszlany, R. Rodriguez-Ferradas, I. Grace, S. W. D. Bailey, K. Gillemot, H. Sadeghi, L. A. Algharagholy, *New J. Phys.* **2014**.
[17] J. M. Soler, E. Artacho, J. D. Gale, A. García, J. Junquera, P. Ordejón, D. Sánchez-Portal, *J. Phys.: Condens. Matter.* **2002**, *14*, 2745.
[18] M. Berritta, D. Z. Manrique, C. J. Lambert, *Nanoscale* **2015**, *7*, 1096-1101.
[19] D. A. Egger, Z.-F. Liu, J. B. Neaton, L. Kronik, *Nano Lett.* **2015**, *15*, 2448-2455.
[20] C. J. Lambert, , *Chemical Society Reviews* **2015** 44, 875-888

# Supplementary Information

# Three States Single-Molecule Naphthalenediimide Switch: Integration of Pendant Redox Unit for Conductance Tuning

Dr. Yonghai Li,[#] Masoud Baghernejad,[#] Al-Galiby Qusiy,[#] Dr. David Zsolt Manrique,[#] Dr. Guanxin Zhang, Joseph Hamill, Dr. Yongchun Fu, PD Dr. Peter Broekmann, Prof. Wenjing Hong,* Prof. Thomas Wandlowski, Prof. Deqing Zhang,* Prof. Colin Lambert*

## 1. Synthesis and characterization of NDI-BT

**NDI-BT** was prepared by Stille-coupling of the respective tin reagent **1** and **2BrNDI** (dibromo-naphthalene diimide, $N$–$C_6H_{13}$, see Scheme S1). **2BrNDI** was synthesized according to the reported procedures.[1] Tin reagent **1** was prepared from its corresponding arylacetylene **2** and used without strict purification, and compound **2** was synthesized according to reported procedures.[2] Other chemicals were purchased from Alfa Aesar, Sigma-Aldrich and used as received without further purification.

$^1$H NMR and $^{13}$C NMR spectra were determined using tetramethylsilane as internal standard at 298 K. Mass spectra were measured in the MALDI-TOF mode. Elemental analysis was performed on a standard elemental analyzer. Cyclic voltammetric measurements were carried out in a conventional three-electrode cell using a glassy carbon working electrode, a Pt counter electrode and a Ag/AgCl(saturated KCl) reference electrode on a computer-controlled CHI660C instruments at room temperature; the scan rate was 100 mV·s$^{-1}$, and $n$-Bu$_4$NPF$_6$ (0.1 M) was used as the supporting electrolyte. Solution and thin films absorption spectra were measured with conventional spectrophotometers.

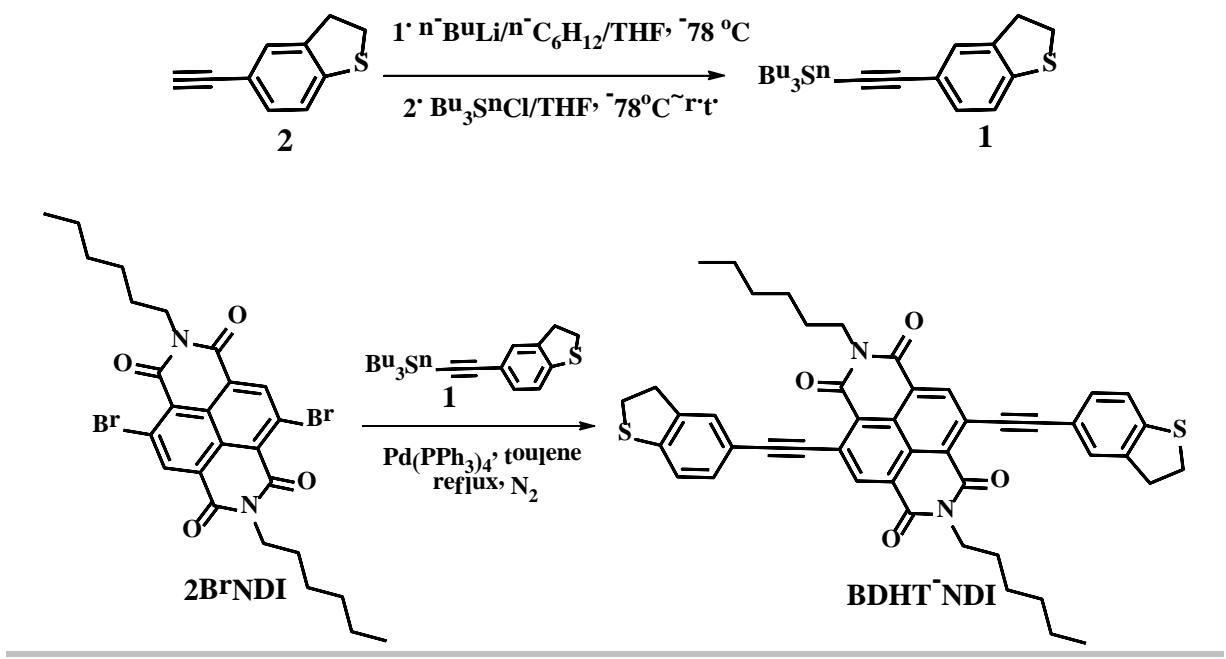

**Scheme S1**. The chemical structure of **NDI-BT** and its synthetic approach.

*Synthesis of 1:* Under $N_2$ atmosphere, to a THF (20 mL) solution of compound **2** (520 mg, 3.25 mmol) 2.45 mL of n-BuLi (1.6 M in n-hexane, 3.9 mmol) was added dropwise at -78 °C. Then, $Bu_3SnCl$ (1 mL, 3.58 mmol) in THF (5 mL) was added to the solution within 30 minutes. The resulting solution was warmed up to room temperature and stirred overnight. The mixture was poured into 20 mL of water and extracted with **n**-hexane (3×20mL). The organic phase was collected, dried with $Mg_2SO_4$ and concentrated to afford **1** as a yellow oil (1.5 g), which was used for the next step without further purification.

*Synthesis of **NDI-BT***: A mixture of 2BrNDI (*N*–$C_6H_{13}$) (100 mg, 0.17 mmol), tin reagent **1** (225mg, 0.50 mmol) and catalytic amount of $Pd(PPh_3)_4$ in toluene (20 mL) was refluxed for 2.5 hours under nitrogen atmosphere. After the reaction, the solvent was evaporated under vacuum and the residue was subjected to silica gel column chromatograph with petroleum ether (60–90 °C)/$CH_2Cl_2$ (v/v, 1/2) as eluent. **NDI-BT** was obtained as a purple solid (115 mg, 0.15 mmol) in 89% yield. $^1$H NMR (400 MHz, $CDCl_2CDCl_2$) δ 8.76 (s, 2H), 7.57 – 7.44 (m, 4H), 7.26 (d, *J* = 8.0 Hz, 2H), 4.25 – 4.10 (m, 4H), 3.42 (dd, *J* = 11.8, 4.3 Hz, 4H), 3.33 (t, *J* = 7.7 Hz, 4H), 1.81 – 1.67 (m, 4H), 1.43 – 1.23 (m, 12H), 0.89 (t, *J* = 6.9 Hz, 6H).$^{13}$C NMR (101 MHz, $CDCl_3$) δ 162.13, 161.63, 145.48, 140.82, 137.19, 132.19, 128.24, 127.11, 126.46, 125.02, 124.86, 122.29, 118.05, 103.87, 89.96, 41.15, 35.86, 33.67, 31.67, 28.08, 26.94, 22.75, 14.24.MS (MALDI-TOF): *m/z* 751.3 (M+1)$^+$. Anal. calcd for $C_{46}H_{42}N_2O_4S_2$: C, 73.57; H, 5.64; N, 3.73; S, 8.54. Found: C, 73.42; H, 5.62; N, 3.71; S, 8.12.

**$^1$H NMR and $^{13}$C NMR spectra:**

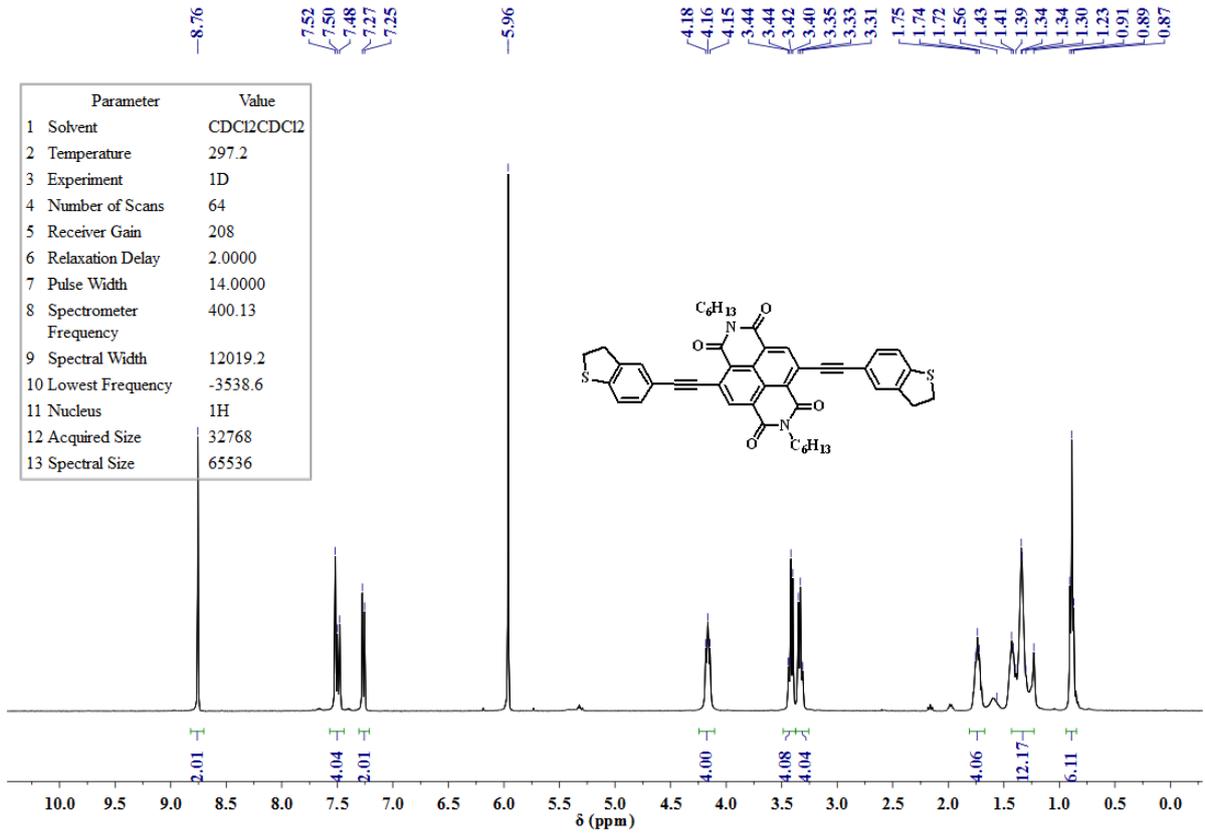

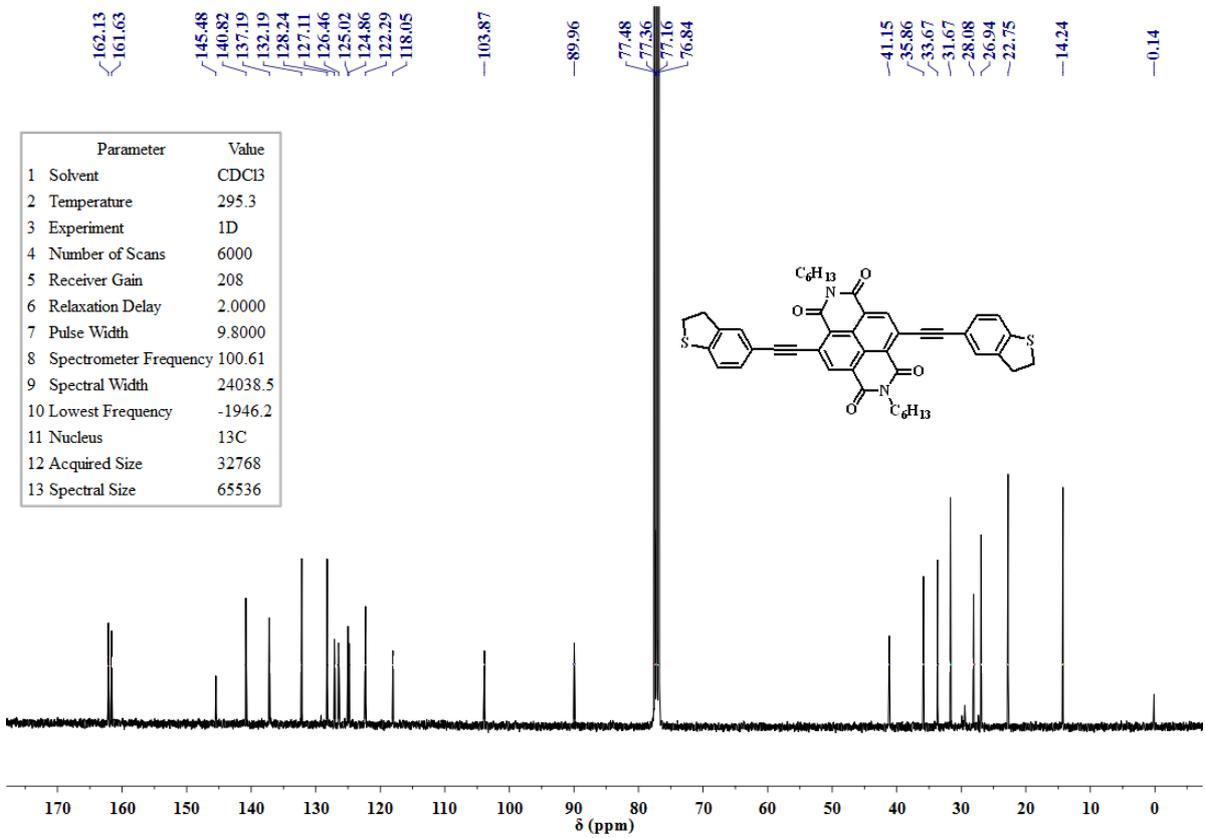

## 2. Spectroscopic and electrochemical studies of NDI-BT

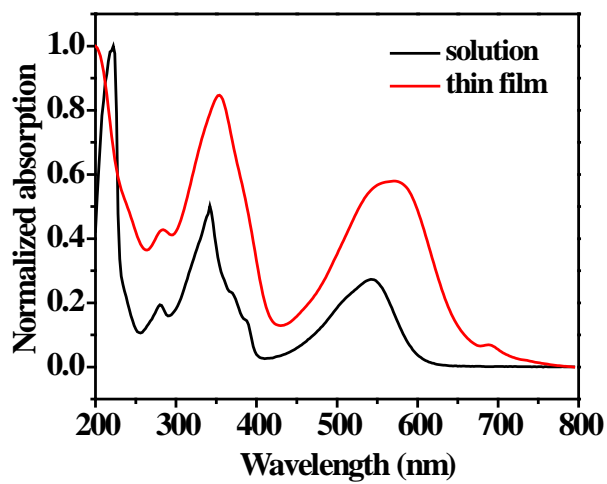

**Figure S1.** The solution (1.0×10$^{-5}$ M, in CH$_2$Cl$_2$) and thin film absorption spectra of **NDI-BT**.

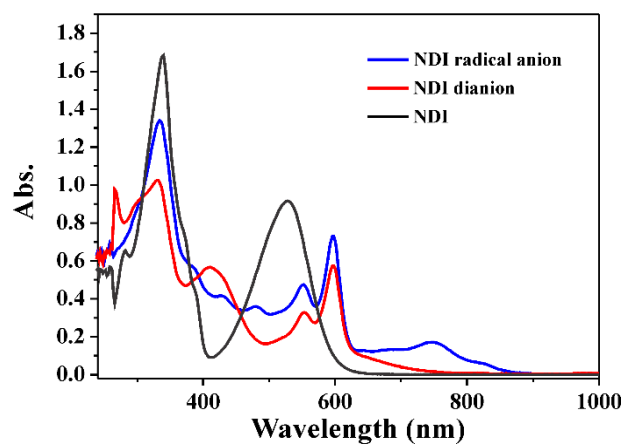

**Figure S2.** UV-Vis spectra of **NDI-BT** neutral, radical, and dianion states in DMF (all concentrations are 20 μM). **NDI-BT** radical and dianion states were obtained by reaction of NDI-BT with pure hydrazine, 20 μL for forming radical anion and 80 μL for di-anion.

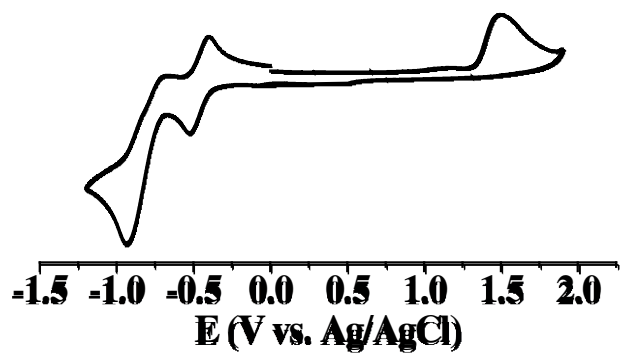

**Figure S3.** Cyclic voltammogram of **NDI-BT** in $CH_2Cl_2$ solutions ($1.0 \times 10^{-3}$ M) at a scan rate of 100mVs$^{-1}$, using a glassy carbon working electrode, a Pt counter electrode and a Ag/AgCl (saturated KCl) reference electrode as the reference electrode, and n-Bu$_4$NPF$_6$ (0.1 M) as supporting electrolyte.

**Table 1**. The absorption data, redox potentials and HOMO/LUMO energies of **NDI-BT**

| | $\lambda_{abs,max}$ (nm)$^a$ | | $E_{red1}^{1/2}$ (V) | $E_{red2}^{1/2}$ (V) | $E_{ox}^{peak}$ (V) | LUMO$^b$ | HOMO$^b$ | $E_g^{cv}$ | $E_g^{opt\ c}$ |
|---|---|---|---|---|---|---|---|---|---|
| | solution | film | | | | | | | |
| **NDI-BT** | 544 | 574 | -0.46 | -0.81 | 1.49 | -4.00 | -5.75 eV | 1.75 eV | 1.83 eV |

$^a$Lowest-energy absorption maxima; $^b$HOMO and LUMO energies of **NDI-BT** were estimated with the following equations: HOMO = -($E_{onset}^{ox1}$ +4.41) eV, LUMO = -($E_{onset}^{red1}$ +4.41) eV; $^c$calculated based on the onset absorption of thin film.

### 3. Single-molecule break junction experiments

The Scanning Tunneling Microscopy Break Junction (STMBJ) experiments were performed with a Molecular Imaging PicoSPM in an environmental chamber and equipped with a dual preamplifier. The current-distance measurements were performed with a lab-build analog ramp unit. The current was recorded at a fixed bias potential during repeated formation and breaking of the molecular junctions. For further technical details we refer to our previous work.[3]

Au(111) was used as the substrate and the facet was cleaned before the experiments using electrochemical polishing and butane flame annealing followed by cooling under Ar atmosphere. Then, the freshly prepared substrate was drop-casted with 30 μL of 0.5 mM **NDI-BT** in THF. For the electrochemically controlled STMBJ experiments, single Pt wires were used as the counter and quasi-reference electrodes and the gold STM tips prepared with AC etching in a 1:1 (v/v) mixture of 30% HCl and ethanol solution, were coated with polyethylene to ensure that the electrochemical current on the tip is below1~2 pA. Finally, the Kel-F cell was mounted on top of the substrate and the supporting electrolyte (ionic liquid HMlmPF6) was added into the cell. The STM images and cyclic voltammograms were recorded frequently during the measurements. The latter were used to check the redox peak position and to ensure that there is no oxygen in the system and no drift of the reference electrode potential. After assembling the experiment, the tip was approached toward the substrate to fulfill the preset tunneling parameters ($i_T$ = 100 pA and a bias voltage $V_{bias}$ = 0.10 V).

After positioning the tip in the tunneling regime, the STM feedback is switched off and current-distance measurements were carried out. For the stretching cycle measurements, the controlling software approaches the STM tip to the drop-casted gold surface. The approach was stopped until a predefined upper current limit was reached to the value corresponds to the formation of several gold-gold contact. After a few ms delay ensuring the formation of stable contacts, the tip was withdrawn until a low current limit of ~10 pA was reached. The approaching and withdrawing rates were both 87 nm/s. The whole current-distance traces were recorded with a digital oscilloscope (Yokogawa DL 750, 16 bit, 1 MHz sampling frequency) in blocks of 186 individual traces. Up to 2000 traces were recorded for each set of experimental conditions to guarantee the statistical significance of the results.

## 4. Theory
### 4.1 Computational Method

To calculate electrical properties for NDI molecule the following method was applied. To begin with, the relaxed geometry of each molecule was found using the density functional (DFT) code SIESTA [4], which employs Troullier-Martins pseudopotentials to represent the potentials of the atomic cores[5], and a local atomic-orbital basis set. In particular, we used the customized basis set definitions to investigate the effects on the Mulliken population count in SIESTA while using the generalized gradient approximation (GGA-PBE) for the exchange and correlation (GGA)[6]. The Hamiltonian and overlap matrix elements are calculated on a real-space grid defined by a plane-wave cutoff of 150 Ry. NDI molecule is relaxed into the optimum geometry until the forces on the atoms are smaller than 0.02 eV/Å. Tolerance of Density Matrix is $10^{-4}$, and in case of the isolated molecules a sufficiently-large unit cell was used. For steric and electrostatic reasons.

After obtaining the relaxed geometry of isolated NDI molecule, we have constructed the junction geometries by placing the optimized NDI molecules between gold electrodes. An example for the junction geometry is shown in Figure 4. After the NDI molecule is placed between the gold electrodes the geometry was again optimized. The optimization were performed with the SIESTA code with the same parameters as used for the isolated molecule. During the relaxation the gold atoms were fixed and for the gold double-zeta basis set were used. The initial distance between S atom (DHBT anchor group) and the centre of the apex atom of each gold pyramid was initially 2.4 Å. After geometry optimization the distance changed to a final value of 2.63 Å.

The charge double layers are added to both sides of the planar backbone of the molecule in the optimized junction geometry as shown in Figures S10-11. The charge double layer is built from sodium and chloride ions with fixed 2.23Å distance between the sodium and chloride ions. After the charge double layer was added to the optimized junction geometry we performed a self-consistent single energy calculation to obtain the optimal electronic structure and performed electron transport calculation with the obtained Hamiltonian and overlap matrices. The charge of the molecule is calculated from the Mulliken population computed by SIESTA. To modify the amount of charge on the molecule we varied the charge double layer-molecular plane distance, denoted Y, while the distance between the ions within the double layer were kept fixed. The distance Y is defined as distance between the plane of the molecule and the center of the closest ion. Figure S5 shows the number of electrons (ΔN) transferred from the NDI molecule to the gold electrodes as a function of distance Y (Å). A few cases with different values of Y are illustrated in Figure S12. To account for the effect of varying environment this calculation was repeated with randomly constructed double layers. The charge double layer is initially constructed as an 2x8 array of sodium ion chloride ion pairs and then four pairs were randomly removed from this array to obtain the randomly constructed double layer.".

The electron and spin transport calculations were performed with the GOLLUM implementation of non-equilibrium Green's function (NEGF) method[7] to compute the transmission coefficient T(E) for electrons of energy E passing from the left gold electrode to the right electrode. GOLLUM is a next-generation code, born out of the SMEAGOL code[8] and uses NEGF combined with density functional theory to compute transport properties of a wide variety of nanostructures. The precise methodology of computing a non-spin-polarized and spin-polarized electron transport over a junction geometry Figure S4 is described in ref[7]. The Hamiltonian and overlap matrices are calculated with SIESTA, using DZP basis sets for all elements except gold, for which DZ basis set was used, GGA-PBE exchange-correlation parameterization, 0.0001 density-matrix tolerance and 200Ry grid cutoff. The Mulliken charges are computed consistently with the same setup.

Once $T(E)$ is computed, we calculate the zero-bias electrical conductance $G$ using the finite temperature Landauer formula:

$$G = I/V = G_0 \int_{-\infty}^{\infty} dE\, T(E)\left(-\frac{df(E)}{dE}\right) \qquad (1)$$

where $G_0 = \left(\frac{2e^2}{h}\right)$ is the quantum of conductance, and $f(E)$ is Fermi distribution function defined as $f(E) = [e^{(E-E_F)/k_B T} + 1]^{-1}$ with $T = 300K$ room temperature[7]. In case of spin-polarized calculation $T(E)$ is the spin-averaged transmission coefficients. Figure S9 compares the zero and room temperature conductances calculated from the same transmission coefficients. Near the resonance peaks the conductance vary with temperature significantly.

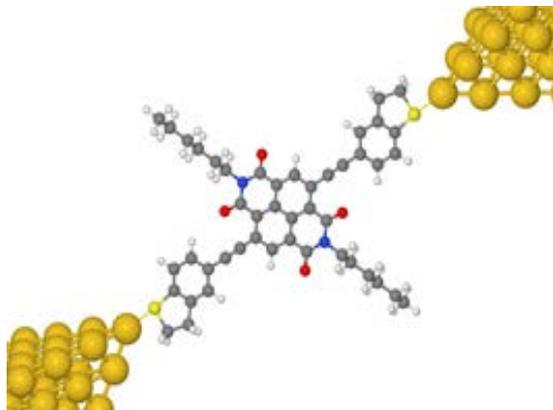

**Figure S4**. Optimized configuration of NDI molecule attached to gold electrodes.

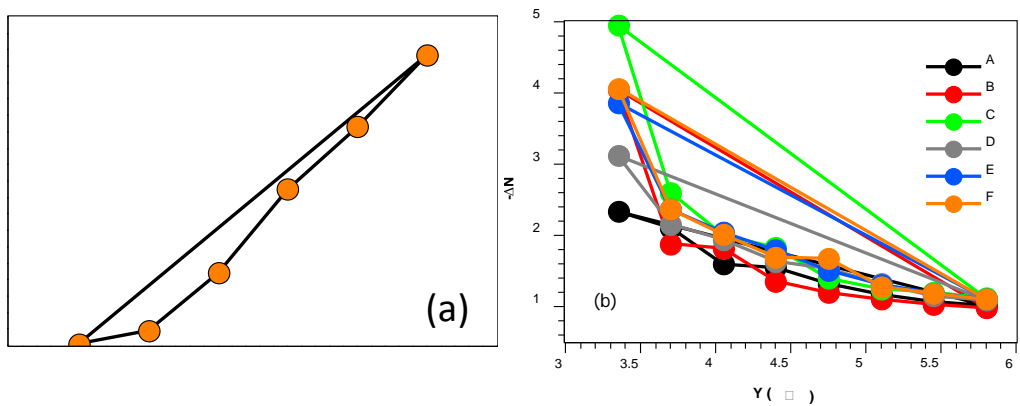

**Figure S5.** The number of electrons (ΔN) transferred from the NDI molecule to the gold electrodes as a function of distance Y (Å) between the molecule and the double layer. (a) shows ΔN when the chloride points towards the NDI, and (b) shows -ΔN when the sodium points towards the NDI (the online color) A, B, C, D, E and F are different randomly-chosen charge double layers.

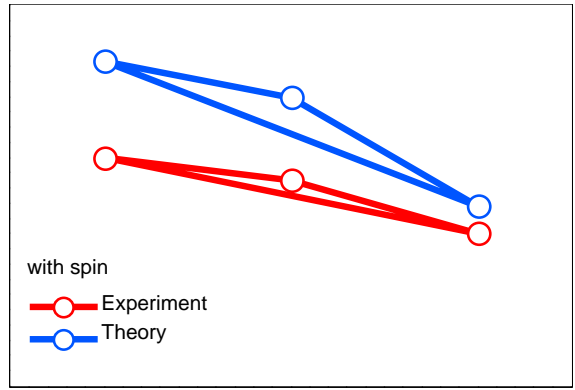

**Figure S6.** DFT-spin polarized calculations of the logarithm scale of electrical conductance of particular configuration of the double layer that show the best comparison between the experimental and theoretical results.

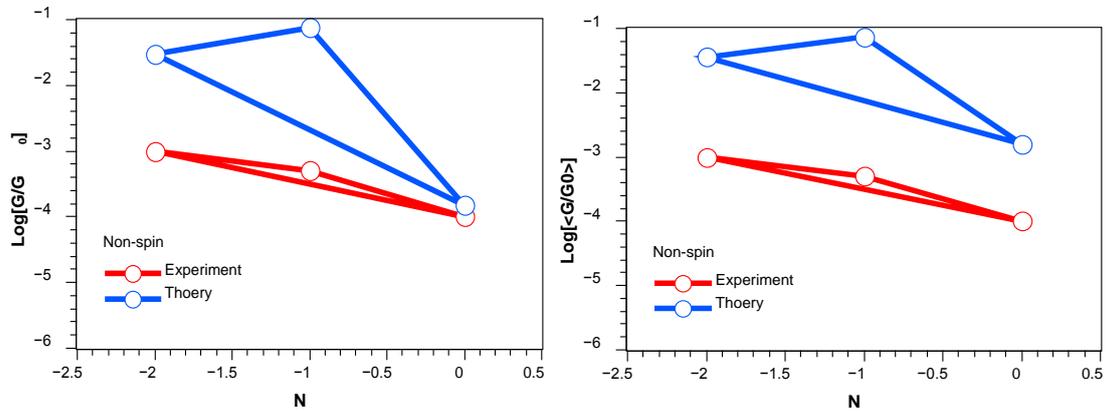

**Figure S7.** DFT Non-spin-polarized calculations to explain the comparison between the experimental and theoretical results, where left shows the logarithm scale of electrical conductance of particular configuration, and right shows the logarithm scale of ensemble averaged of electrical conductance of randomized configurations.

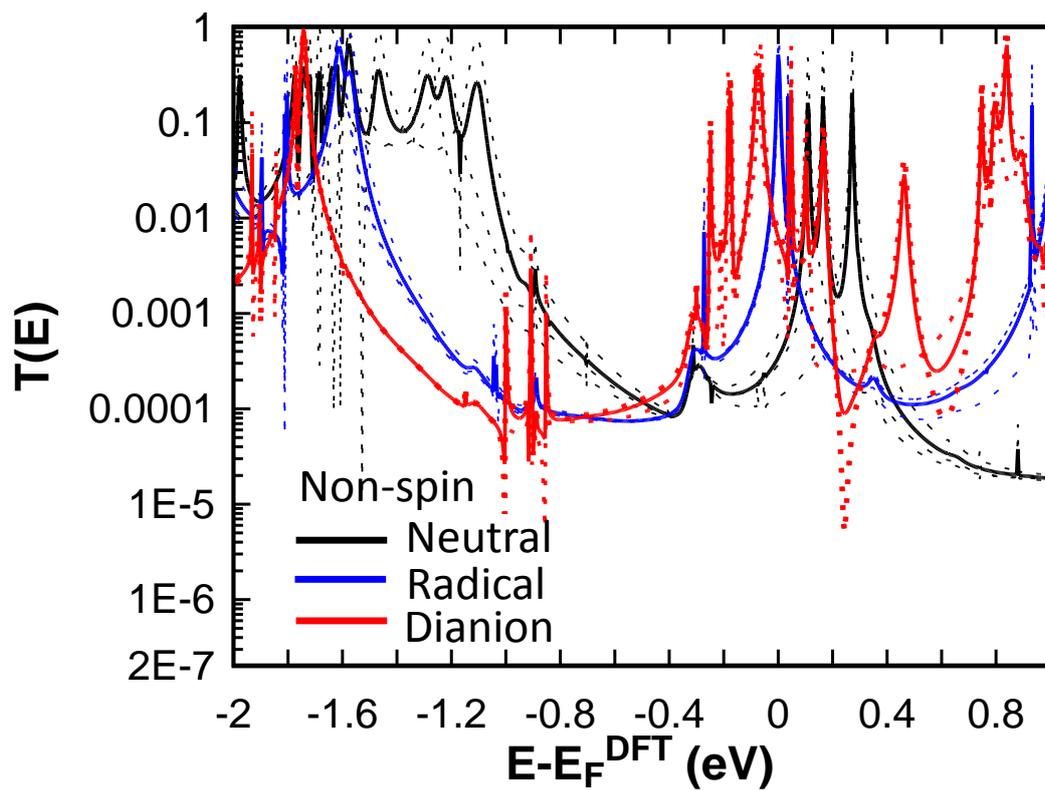

**Figure S8.** shows (the dotted lines) the transmission coefficient curves for randomized configurations of double layers with non-spin polarization, where the thick lines are the averaged transmission coefficient as a function of energy for NDI-molecule attached to the electrodes in the three cases: neutral (black), radical (blue), and dianion (red).

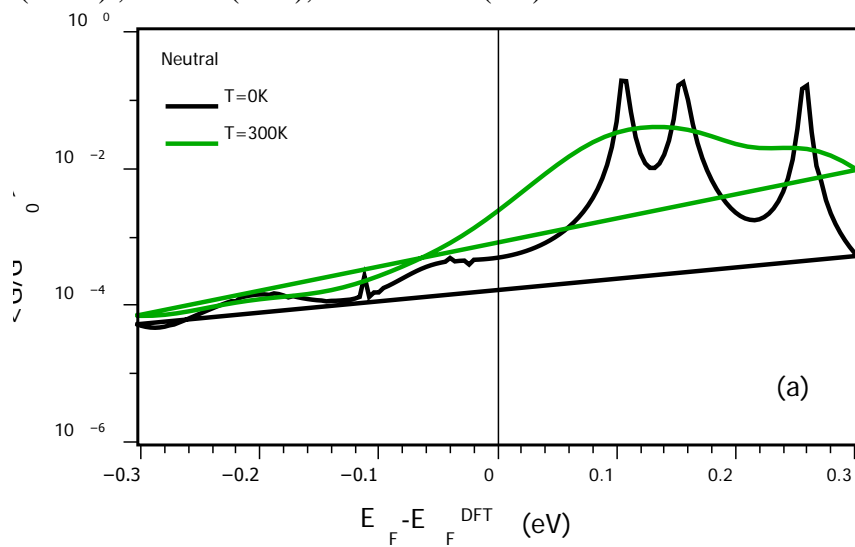

(a)

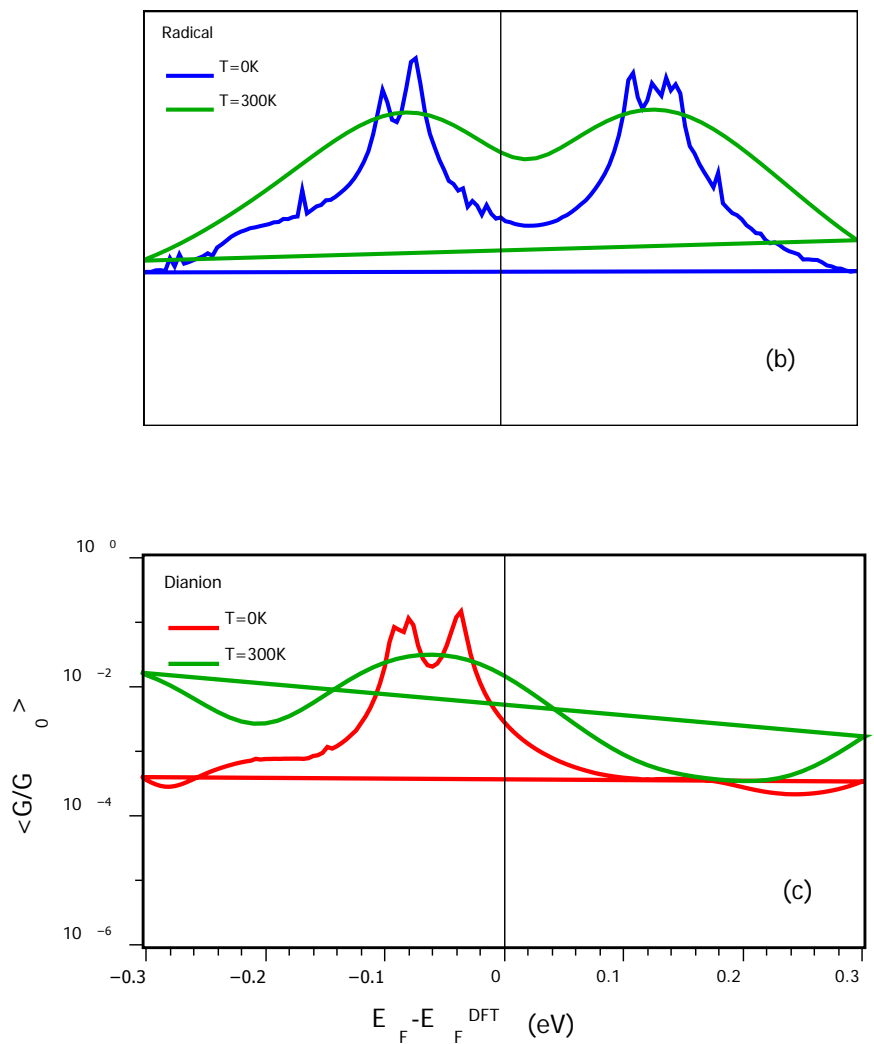

**Figure S9.** Shows the comparison between the room temperature and zero temperature conductances for a range of Fermi energies.

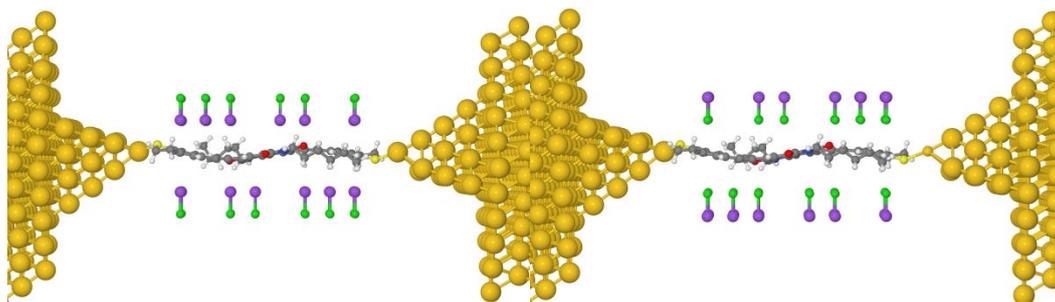

**Figure S10.** NDI molecule attached to gold electrodes in the presence of charge double layer above and below the molecule. The positive sodium ion is purple and the negative chloride ion is green.

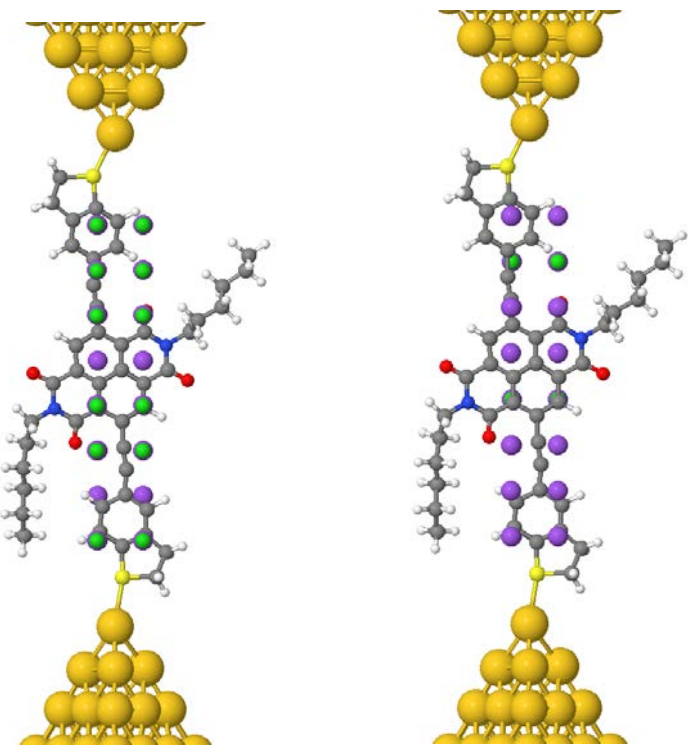

**Figure S11.** shows the geometries on Figure S10 from another orientation to illustrate the charge double layer around the backbone of the molecule.

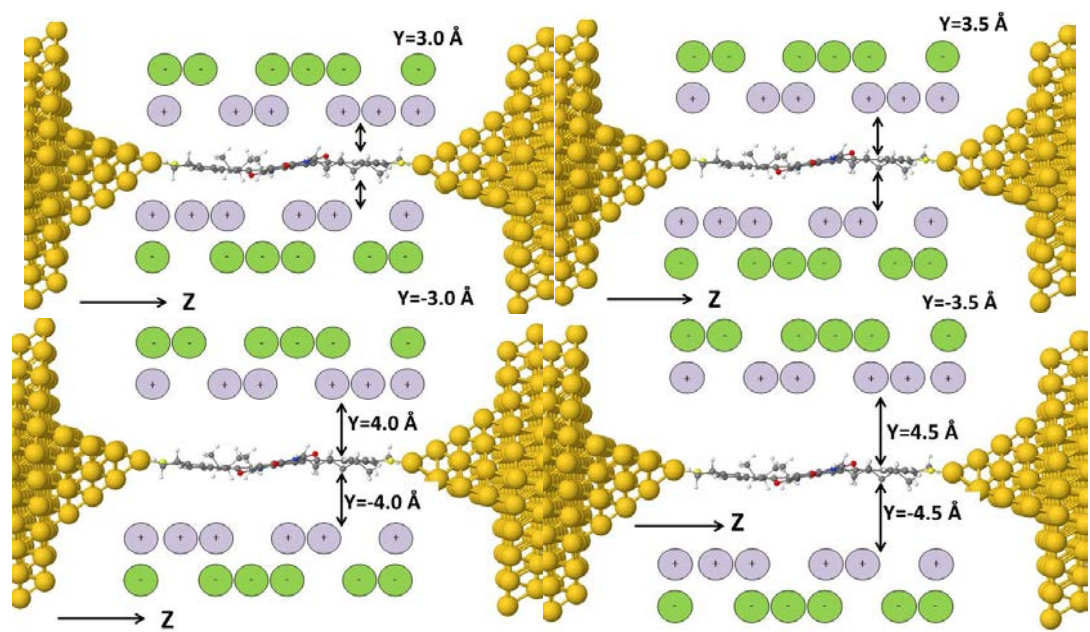

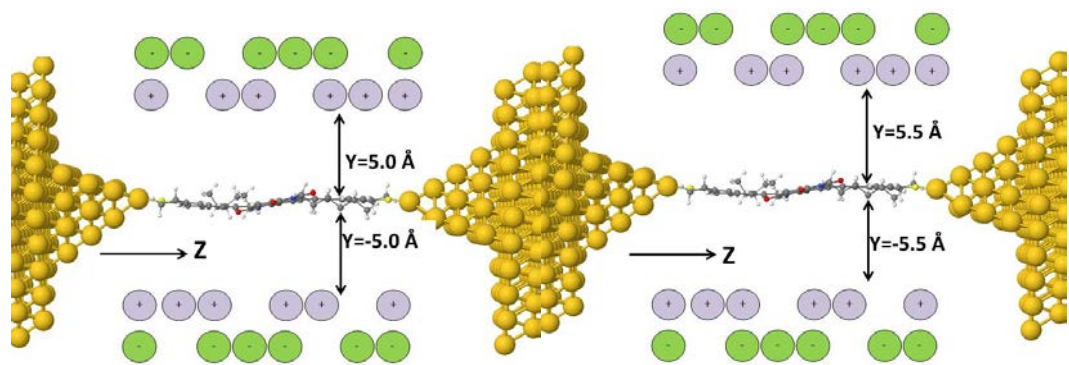

**Figure S12.** Illustrates charge double layer with varying Y distances.

# Reference

[1] F. Chaignon, M. Falkenstrom, S. Karlsson, E. Blart, F. Odobel, L. Hammarstrom, *Chemical Communications* **2007**, 64-66.
[2] a) P. Moreno-Garcia, M. Gulcur, D. Z. Manrique, T. Pope, W. Hong, V. Kaliginedi, C. Huang, A. S. Batsanov, M. R. Bryce, C. Lambert, T. Wandlowski, *J. Am. Chem. Soc.* **2013**, *135*, 12228-12240; b) D. Madec, F. Mingoia, C. Macovei, G. Maitro, G. Giambastiani, G. Poli, *European Journal of Organic Chemistry* **2005**, 552-557.
[3] C. Li, I. Pobelov, T. Wandlowski, A. Bagrets, A. Arnold, F. Evers, *Journal of the American Chemical Society* **2008**, *130*, 318-326.
[4] J. M. Soler, E. Artacho, J. D. Gale, A. García, J. Junquera, P. Ordejón, D. Sánchez-Portal, *J. Phys.: Condens. Matter.* **2002**, *14*, 2745.
[5] N. Troullier, J. L. Martins, *Physical Review B* **1991**, *43*, 1993.
[6] a) J. P. Perdew, K. Burke, M. Ernzerhof, *Physical review letters* **1996**, *77*, 3865; b) B. Hammer, L. B. Hansen, J. K. Nørskov, *Physical Review B* **1999**, *59*, 7413.
[7] J. Ferrer, C. Lambert, V. García-Suárez, D. Z. Manrique, D. Visontai, L. Oroszlany, R. Rodríguez-Ferradás, I. Grace, S. Bailey, K. Gillemot, *New Journal of Physics* **2014**, *16*, 093029.
[8] A. R. Rocha, V. M. Garcia-Suarez, S. W. Bailey, C. J. Lambert, J. Ferrer, S. Sanvito, *Nature materials* **2005**, *4*, 335-339.
[9] X.H. Zheng, G.R. Zhang, Z. Zeng, V.M. García-Suárez, C.J. Lambert, (2009) Phys. Rev B 80 (7), 075413
[10] S.W.D. Bailey, I. Amanatidis, C.J. Lambert, (2008) Phys, Rev. Lett. 100 (25), 256802